\documentclass[prb,twocolumn,superscriptaddress,showpacs]{revtex4}
\usepackage[italian,english]{babel}

\usepackage{graphicx}
\newcommand{\be}{\begin{equation}}
\newcommand{\ee}{\end{equation}}
\newcommand{\bea}{\begin{eqnarray}}
\newcommand{\eea}{\end{eqnarray}}
\newcommand{\bd}{\begin{displaymath}}
\newcommand{\ed}{\end{displaymath}}
\renewcommand{\a}{\alpha}
\renewcommand{\b}{\beta}
\renewcommand{\r}{\rho}
\newcommand{\s}{\sigma}
\newcommand{\g}{\gamma}

\renewcommand{\l}{\lambda}
\renewcommand{\d}{\delta}
\newcommand{\hf}{\frac{1}{2}}
\newcommand{\F}{\Phi}

\renewcommand{\o}{\omega}
\newcommand{\lf}{\left}
\newcommand{\rg}{\right}
\renewcommand{\O}{\Omega}

\begin{document}

\title{Qubits as devices to detect the third moment of current fluctuations}

\author{Valentina Brosco}
\affiliation{NEST-CNR-INFM \& Dipartimento di Fisica, Universit\'a di Pisa,
    largo E. Fermi, I-56100 Pisa, Italy}
\author{Rosario Fazio}
\affiliation{NEST-CNR-INFM \& Scuola Normale Superiore, piazza dei
        Cavalieri 7, I-56126 Pisa, Italy}
\affiliation{International School for Advanced Studies (SISSA)
        via  Beirut 2-4,  I-34014, Trieste - Italy}
\author{F.W.J. Hekking}
\affiliation{Laboratoire de Physique et Mod\'elisation des Milieux
        Condens\'es Magist\`ere-CNRS, B.P. 166, 38042 Grenoble cedex 9, France}
\author{J.P. Pekola}
\affiliation{Low Temperature Laboratory, Helsinki University of
        Technology, P.O. Box 3500, FIN-02015 HUT, Finland}

\begin{abstract}
Under appropriate conditions controllable two-level systems can be
used to detect the  third moment of current fluctuations. We
derive a Master Equation for a quantum system coupled to a bath
valid to the third order in the coupling between the system and
the environment. In this approximation the reduced dynamics of the
quantum system depends on the frequency dependent third moment.
Specializing to the case of a controllable two-level system (a
qubit) and in the limit in which the splitting between the levels
is much smaller than the characteristic frequency of the third
moment, it is possible to show that the decay of the qubit has
additional oscillations whose amplitude  is directly proportional
to the value of the third moment. We discuss an experimental
setup where this effect can be seen.
\end{abstract}


\maketitle

\section{Introduction}

A comprehensive understanding of the transport properties of
mesoscopic conductors can be achieved with the study of both the
average current and its fluctuations. The investigation of shot
noise~\cite{deJong97,blanter99,Kogan96} has proven to be a valuable
tool to determine properties which are elusive to the study of current-voltage
characteristics. One of the most celebrated examples in this respect
is probably the measurement of the fractional charge by means of the study of
shot-noise in point contacts in the Fractional Quantum Hall
regime~\cite{depicciotto97,saminadayar97}.

In the case of non-gaussian fluctuations, moments beyond the
second one are relevant in characterizing the transport. In the
last few years numerous theoretical studies (see
Ref.\onlinecite{delft}) analyzed the properties of higher moments
and of the full counting statistics~\cite{levitov96}. In contrast
to the large theoretical activity, experiments are very difficult
and only few appeared so far.  The first pioneering measurement of
the third moment, performed by Reulet {\em et al}~\cite{reulet03},
has been hindered by environmental effects~\cite{beenakker03}. The
first three moments of the current fluctuations in a tunnel
barrier were measured very recently by Bomze {\em et
al.}~\cite{bomze05} confirming the Poisson statistics associated
with the discreteness of the charge. Further experimental
indications on the non-Gaussian character of noise were obtained
by Lindell {\em et al.}~\cite{lindell04} who observed its effects
on a Coulomb blockade Josephson junction.

 In parallel with the first experiments,
and with the hope of finding more effective ways to measure higher
moments, several theoretical papers appeared suggesting ways to
find signatures of non-gaussian noise in the non-equilibrium
properties of mesoscopic systems used as detectors. The
first practical way  to probe high moments of current  was
suggested by Lesovik in Ref.\onlinecite{lesovik94}. In
Ref.\onlinecite{aguado}, Aguado and Kouwenhoven  considered the
possibility to use a double quantum dot system as detector of high
frequency noise. More recently, Josephson junctions were shown to
be able to act as detectors of the third~\cite{heikkila04} and
fourth moments of current fluctuations\cite{ankerhold05}. Their
use as threshold detectors to measure the full counting statistics
has been discussed by Tobiska and Nazarov~\cite{tobinska04}, by
Lesovik, Hassler and Blatter~\cite{lesovik06} and by one of the
authors~\cite{pekola04}.

Qubits have been already proven very sensitive spectrometers of
noise~\cite{devoret00, astafiev04}. In this work we want to
explore further the use of controllable two-level systems as noise
spectrometers and analyze the possibility to employ them for the
measurement of the third moment as well. With this scope in mind
we derive a perturbative equation for the dynamics of two-level
systems in presence of noise up to the third order in the
system-noise coupling to see if, under some circumstances, we can
extract some information on the third moment. In general the third
order effects are masked by the dominant second order ones, since
they are a result of a perturbative expansion. There are, however,
situations in which the second-order correction vanishes and
therefore the third order is the leading  contribution.
 We will show that, in the
usual Rotating Wave Approximation (RWA) of the system equations of
motion, the contribution  of the third moment is a small
correction to the dominant effect of the second moment (and hence
difficult to measure). A treatment beyond RWA is therefore needed
and it leads to the presence of additional effects solely due to
the third moment of current fluctuations.

The paper is organized as follows. In the next Section we introduce
the model. We then proceed  to derive the Master
Equation for the reduced dynamics of the quantum system up to third order
in the coupling with the environment.
The reduced dynamics of the quantum system will also depend on the
third-order correlations of noise and therefore it may act as a
detector of these higher order moments. In
Section~\ref{relaxation} we concentrate on the case in which the
quantum system is a two-level system and show that the presence of
the third order may induce coherent oscillations in the ground
state population of the quantum system. Furthermore, we show that
the amplitude of these oscillations is proportional to the
three-point correlator of the fluctuations. In Section \ref{mw} we
discuss the case where the two-level system is subject to an
external microwave field. In particular we discuss how Rabi
oscillations can be influenced by the presence of the third order
noise. The motivation here is to lower the frequency of the
coherent oscillations into a regime which would be more accessible
to experiments. Possible experimental setups where these effects
can be measured are discussed in Section~\ref{exps}. In the same
Section we analyze various complications which may emerge in the
actual measurement. In particular we consider the case of a
DC-SQUID as a third-moment detector. Section~\ref{conclusions} is
devoted to a summary of the results and possible perspectives of
this approach in measuring higher order current fluctuations.
Recently  a similar detection scheme has been discussed in
Ref.\onlinecite{ojanen05}; our approach is different in spirit and
we will point out the difference with Ref. \onlinecite{ojanen05}
where the Rotating Wave approximation is taken for
granted~\cite{falci}.

\section{Dynamics of the qubit}
\label{master}

\subsection{The model}
In this first section we discuss the general formalism which will be
used in the remainder of the paper. The setup under consideration is
composed of a quantum system $S$ weakly interacting with a quantum
bath $B$. As explained in the Introduction the quantum system will
be used to investigate the properties of the bath which for example
may be another nanostructure (a tunnel barrier, a point contact,...)
biased at a fixed voltage and of which we want to study current
fluctuations. The Hamiltonian of the total system $S+B$ can be
written as follows:
\begin{equation}
    \hat{H}_T = H_S+H_B+\hat{V},
\end{equation}
where $H_S$ and $H_B$ are respectively the free Hamiltonian of
 the system and of the bath. The interaction potential, $\hat{V}$,
is chosen to be of the form
\begin{equation}
    \hat{V} = g\sum_{\a}\hat{N}_\a\otimes \hat{Q}_\a.
\end{equation}
In the definition of $\hat V$, $g$ is an adimensional coupling
constant and $\hat{N}_\a$ and $\hat{Q}_\a$ are operators of the
bath and system respectively. The interaction is chosen to be weak
so that the dynamics of the reduced density matrix of the system
can be obtained by a perturbation expansion in $g$. The procedure
is well known and described in various textbooks~\cite{books}. It
is typically performed up to second order in the coupling $g$;
here we do a step forward and go to the next order in the
coupling. As we focus our attention on the study of the time
evolution of the system $S$ in presence of a stationary bath, we
have $[\r_B,H_B]=0$,  $\r_B$ being the bath density matrix.
Moreover we assume  the dynamics of the whole system to be
Markovian. This means that at each order of perturbation theory we
can neglect all the terms that are non-local in time.

\subsection{Third order Master Equation}
The time evolution of the reduced density matrix of the system in the
interaction representation is described by the following third
order equation (the steps leading to the master equation are
standard~\cite{books} and we do not repeat them):
\begin{widetext}
\be
\label{bloch}
    \dot{\r_I}=Tr_B\left\{-\frac{g^2}{\hbar^2}\int_0^\infty\!\!
    dt'\big[V(t),\left[V(t'),\rho_I(t)\rho_B\right]\big]+\frac{ig^3}{\hbar^3}\int_0^\infty\!\!
    dt'\int_0^{t'} \!\!dt''
    \Big[V(t),\big[V(t'),\left[V(t''),\rho_I(t)\rho_B\right]\big]\Big]\right\},
\ee
\end{widetext}
where we denoted respectively with $\r_I$ and $V(t)$, the density
matrix of the system and the interaction potential in the
interaction representation. In deriving Eq.(\ref{bloch}) we made
the further assumption that $\lf<V\rg>=Tr_B[\r_BV]=0$, where
$Tr_B$ denotes the trace over the bath degrees of freedom. Taking
the matrix elements of (\ref{bloch}) between two eigenstates of
the Hamiltonian of the system, after some algebra, we obtain the
following third order master equation for the density matrix of
$S$: \be \label{m.e.}
    \dot\rho_{I,mn}=
    \sum_{kl}\rho_{I,lk}D_{knml} e^{i(E_m-E_n-E_k+E_l)t},
\ee
where we set $H_S\lf.|m\rg>=E_m\lf.|m\rg>$,
$\r_{I,mn}=\lf<m|\r_I|n\rg>$ and the third order relaxation matrix,
$D_{knml}$, is given by the sum of two contributions:
\be
\label{rm}
        D_{knml}=\frac{g^2}{\hbar^2}R_{knml}-\frac{i
        g^3}{\hbar^3}C_{knml} \;\; .
\ee
In the previous equation $R_{knml}$ is the second order
relaxation matrix and $C_{knml}$ is a third order
correction crucial to our treatment. Here we omit to show the explicit expression of
 $R_{knml}$  as it is well known and it can be found in  textbooks \cite{books}.\\
The third order kernel
$C_{knml}$ can be written as follows: \be
    C_{knml}=A_{knml}-A^*_{lmnk}+B_{mlkn}-B^*_{nklm},
\ee with
\begin{widetext}
\bea
    A_{knml}& =&\sum_{\a\b\g} \sum_i\left[\sum_j
    Q^\a_{mi}Q^\b_{ij}Q^\g_{jl}\d_{kn}F_{\a\b\g}(\o_{li},\o_{ij})-
    Q^\b_{mi}Q^\g_{il}Q^\a_{kn}F_{\a\b\g}(\o_{lm},\o_{mi})\right]\label{Aknml}\\
    B_{mlkn}& =& \sum_{\a\b\g}\sum_i\left[
    Q^\b_{ml}Q^\g_{ki}Q^\a_{in}G_{\a\b\g}(\o_{lm}+\o_{ik},\o_{ml})-
    Q^\a_{mi}Q^\b_{il}Q^\g_{kn}G_{\a\b\g}(\o_{li}+\o_{nk},\o_{il})\right]\label{Bknml}
\eea
and where $\hbar \o_{mn}=E_m-E_n$. The functions
$F_{\a\b\g}$, $G_{\a\b\g}$ are the three-point correlators of the
noise operators:
\bea
    F_{\a\b\g}(\o_1,\o_2)&=&\int^\infty_0dt_1\int_0^{t_1}dt_2
    \left<N_\a(t_1)N_\b(t_2)N_\g(0)\right>_Be^{i\o_1t_1}e^{i\o_2t_2}
\label{F}\\
    G_{\a\b\g}(\o_1,\o_2)&=&\int^\infty_0dt_1\int_0^{t_1}dt_2\left<N_
    \g(0)N_\a(t_1)N_\b(t_2)\right>_Be^{i\o_1t_1}e^{i\o_2t_2} \;\; .
\label{G}
\eea
\end{widetext}
 We used
$Q^\a_{mn}$ instead of $\left<m\,\vline\, \hat Q_\a\,\vline\,
n\right>$, to denote the matrix elements of a system operator in the
Schr\"odinger picture.\\
The average $\lf<...\rg>_B \equiv Tr_B[\r_B ...]$ is taken over the
density matrix of the bath. Note that we do not need the bath
to be in equilibrium but we do assume that it is stationary. An example
is the noise generated by a non-equilibrium current in a voltage
biased tunnel junction.\\

Equation (\ref{m.e.}) is quite general and many specific cases can
be studied starting from it. In the Rotating Wave Approximation,
i.e. neglecting oscillating terms in the sum on the right-hand side
of Eq.(\ref{m.e.}), one can recover the result of
Ref.~\onlinecite{ojanen05}. In this case the presence of the third
order causes simply a small correction of the second order
transition amplitudes and therefore it might be difficult to detect
in the presence of a large background due to the second order
contribution. In the following sections, we will analyze in more
detail some special cases in which the effects of the third order
relaxation matrix can be well characterized and distinguished from
the second order. In particular, we will study how the third order
contribution affects the decay and the Rabi oscillations of a
two-level probe quantum system.

\section{Results}
\label{results}
We now specialize to the case in which the probe
is a two-level quantum system. We assume that the
effective Hamiltonian of the system in the presence of noise
has the form
\be
\label{heff}
    \hat H_{eff}= -\hbar\o_0\hat \sigma_z+N_T(t)\hat \sigma_x
\ee when expressed in the eigenbasis of $S$, $N_T(t)$ is the noise
operator and $\hat \sigma_{i}$ are the Pauli matrices. Moreover we
make the hypothesis that the relevant frequencies of the noise
source are much larger than the level splitting  $\o_0$. We thus
neglect the frequency dependence of the third order correlators on
scale up to  $\o_0$. Consequently in the calculation of the third
order coefficients of the relaxation matrix,  we set
$F(\o_1,\o_2)\simeq F(0,0)$ and $G(\o_1,\o_2)\simeq G(0,0)$, if
$\o_1,\o_2\sim \o_0$. In the following we will comment on these
assumptions.

\subsection{Relaxation in presence of non-gaussian noise}
\label{relaxation}
In case of a two-level system the third order master equation, Eq.(\ref{m.e.}), in the Schr\"odinger representation, reduces to
\begin{widetext}
\bea
\label{sch}
    \dot\r_{11}&=&\lf(D_{1111}-D_{2112}\rg)\r_{11}+\lf(D_{1112}+D_{2111}\rg)\textrm{Re}
    [\r_{12}]-i\lf(D_{1112}-D_{2111}\rg)\textrm{Im}[\r_{12}]+D_{2112}
    \\ \label{sch1}
    \dot\r_{12}&=&\lf(D_{1211}-D_{2212}\rg)\r_{11}+D_{2212}
    -i\lf(D_{1212}-D_{2211}-i\o_0\rg)\textrm{Im}[\r_{12}]+\lf(D_{2211}+
    D_{1212}+i\o_0\rg)\textrm{Re}[\r_{12}].
\eea
\end{widetext}
where $\hbar\o_0=E_2-E_1$. The different elements of the third order relaxation
matrix, $D_{knml}$, can be calculated using the definition given in
the previous paragraph, Eqs. (\ref{rm})-(\ref{G}). Within our
hypothesis, the only nonzero second order contributions are
\bea
    D_{2112}&=&\frac{g^2}{\hbar^2}
    \int^\infty_{-\infty}dt'\left<N_T(t')N_T(0)\right>e^{i\o_0t'}=W_{21}\label{w21}\\
    D_{1111}&=& -D_{2112}(-\o_0)=-W_{12}\label{w12}\\
    D_{1212}&=&\frac{g^2}{\hbar^2}\int^\infty_{0}dt'\left<\{N_T(t'),N_T(0)\}
        \right>e^{i\o_0t'}\\\label{deph}
    D_{2211}&=&-D^*_{1212}.
\eea where we have introduced the second order transition rates,
$W_{12}$ and $W_{21}$.\\ Note that, due to our transverse coupling
assumption,the third order contribution to the previous matrix
elements, Eqs. (\ref{w21})-(\ref{deph}), is zero. The other
coefficients of the relaxation matrix are of the third order in
the coupling constant, $g$. In the limit of a flat spectrum all
these elements can be defined as follows, using only one
independent parameter \bea
    \!\!\!\!\!D_{1112}\!=\!D_{2212}=-i\Lambda^{(3)}\,\,\,\,\,
    D_{2111}\!=\!D_{1211}=i\Lambda^{(3)}
\eea The third order coefficient, $\Lambda^{(3)}$, is real and it
can be written a sum of time-ordered products as follows

\bea\label{lambda3}\Lambda^{(3)}&\!\! =&\!\!
\frac{g^3}{\hbar^3}\int\! dt_1\!\int\! dt_2\Big[\lf<N_T(t_1) T
\lf(N_T(t_2)N_T(0)\rg)\rg>+\nonumber\\& & +\frac{1}{3}\lf<\tilde T
\lf(N_T(t_1)N_T(t_2)N_T(0)\rg)\rg>\Big] \eea

where $T$ and $\tilde T$ denote respectively the time-ordering and
the anti-time-ordering operator.\\

Third moment fluctuations can be measured by measuring the probability that the
system is in the ground state once it was initially prepared in the state
$$
    \lf |\Psi(t=0)\rg>=\frac{1}{\sqrt{2}}\lf(\lf |1\rg>+\lf |2\rg>\rg) \, .
$$
Indeed the ground state population as a function of time can be easily
calculated from the integration of Eqs.(\ref{sch})-(\ref{sch1})
\be
\label{solution}
    \r_{11}(t)=\frac{W_{21}}{W_{12}+W_{21}}+e^{-(W_{12}+W_{21})
    t}\lf[A+2B\cos(\o_R t)\rg].
\ee
In the previous equation, we have introduced the renormalized frequency
\be
    \o_R^2=\o_0^2+\o_0\textrm{Im}\lf[D_{1212}\rg]-\frac{1}{4}(W_{12}+W_{21})^2
\ee and the coefficients $A$ and $B$ which are defined by the
following equations
\bea
    A&=&\hf\frac{W_{12}-W_{21}}{W_{12}+W_{21}}-\frac{\Lambda^{(3)}}{2\o_R}+O(g^4)\\
    B&=&\frac{\Lambda^{(3)}}{4\o_R}+O(g^4).\label{B}
\eea
\begin{figure}
\begin{center}
\includegraphics[scale=0.75]{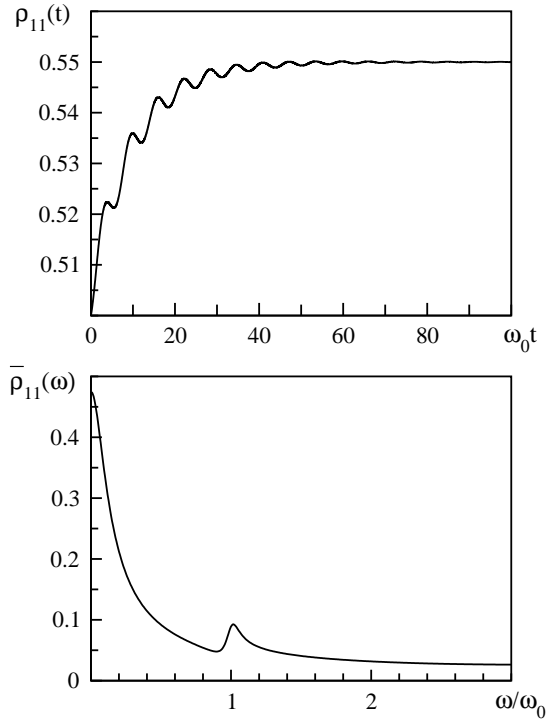}

\caption{Ground state population as a function of time and its
    Fourier transform calculated for the following values of the
    parameters: $W_{12}+W_{21}=\o_0g^2$,
    $W_{12}-W_{21}=0.05\,\o_0g^2$, $\Lambda^{(3)}\sim 0.3\,\o_0g^3$, $\o_R=
    \o_0$, $g^2= 0.1$ .}
\label{senza}
\end{center}
\end{figure}

The presence of the third order, or more precisely, the presence of
nonzero odd moments of noise fluctuations, induces measurable
effects in the dynamics of the probe quantum system. As one can see from
Eqs.(\ref{solution})-(\ref{B}), it induces coherent oscillations in
the ground state population of amplitude proportional to the third
order parameter, $\Lambda^{(3)}$.

In Fig.\ref{senza} we show the ground state population as a
function of time and its Fourier transform. As one can see
$\r_{11}(t)$ is given by the superposition of two terms: a damped
exponential whose asymptotic value is fixed by the ratio between
the two transition amplitudes, $W_{12}$ and $W_{21}$, and a damped
cosine term proportional to the third moment. The structure of
$\r_{11}(t)$ can be analyzed by studying its Fourier transform
(shown in Fig.\ref{senza} lower panel) defined as
$$
    \bar f(\o)=\mid \int_{-\infty}^\infty dt\, e^{i\o
    t}\lf[f(t)-f(t\!=\!\infty)\rg] \mid \,\, .
$$
The zero frequency peak is related to the second order
non-oscillating contribution, while the  smaller peak at frequency
$\o_0$ is a pure third order effect. In absence of third moment
the time dependence of the ground state population would be simply
described by a damped exponential and no third order peak would
appear in the Fourier transform at $\omega = \omega _0$.

The assumption that the noise couples to the system only through $\sigma_x$
(transverse coupling) was crucial in our analysis  to
separate the second and the third order contributions in different
elements of the relaxation matrix. This assumption can be  relaxed, by
introducing in the Hamiltonian a longitudinal  term of the form: $V_L(t)=g_L
N_L(t)\sigma_z$, provided that the two noise operators $N_T(t)$ and
$N_L(t)$ can be considered as uncorrelated and that $V_L(t)$ is
weak. In this case the final result is essentially the same except
for a redefinition of order $g_L^2$ of the transition amplitudes and
of the renormalized frequency, $\o_R$.

\subsection{Effects of a microwave field}
\label{mw} As shown in the previous section the presence of odd
moments in the current fluctuations has a distinct signature in
the oscillations of the ground state population in the case of
transverse coupling to the noise, Eq.(\ref{heff}).

 However, the
actual measurement of these oscillations can be very difficult as
their characteristic frequency, $\o_R \simeq \o_0$, is typically
of the order of 10 GHz and the time resolution required to follow
such oscillations in detail is hardly accessible. In this Section
we discuss a generalization of the the case discussed  before to
account for the presence of an external microwave field. Our aim
is to clarify under which conditions a microwave field can shift
the third order peak to a lower value fixed by the detuning
frequency.

In presence of microwaves,  the dynamics of the quantum system can be described
by the effective hamiltonian: $ \hat H_{eff}=H_S+\hat{V}+\hat{M}$.
The effect of the applied field leads to the term $
  \hat{M}=\hat{O}\cos(\O t)\label{hatM}
$ where $\hat{O}$ is a system operator which, quite generally, can be expressed
in the form
\be
\label{op}
    \hat{O}=\frac{\hbar}{2}\lf(M_L\hat \s_z+M_T\hat \s_x\rg) \, .
\ee

The corresponding Master Equation for a two-level quantum system
in the presence of a microwave field is
\begin{widetext}
\bea
    \dot{\rho}_{11}&=&\lf(D_{1111}-D_{2112}\rg)\r_{11}+
    \lf(D_{1112}+D_{2111}\rg)\textrm{Re}[\r_{12}]+D_{2112}
    -i\big[D_{1112}-D_{2111}-iM_T\cos(\O t)\big]\textrm{Im}[\r_{12}]
    \nonumber\\ & &\label{mws1}\\
    \dot{\rho}_{12}&=& \lf[D_{1211}-D_{2212}+iM_T\cos(\O t)\rg]\r_{11}+
    \lf[M_L\cos(\O t)-\o_0-i\lf(D_{1212}-D_{2211}\rg)\rg]\textrm{Im}[\r_{12}]
    \nonumber\\
    & & +\lf[D_{2211}+D_{1212}-i\lf(M_L\cos(\O t)-\o_0\rg)\rg]\textrm{Re}
    [\r_{12}]+D_{2212}-iM_T\cos(\O t).\label{mws2}
\eea
\end{widetext}

\begin{figure}
\begin{center}
\includegraphics[scale=0.75]{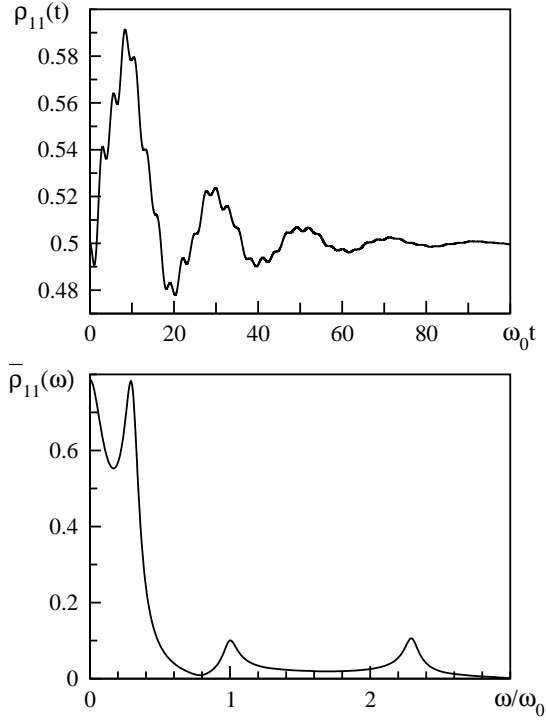}
\caption{Ground state population as a function of
    time (upper panel) and its Fourier transform (lower panel)
    calculated in presence of a weak transverse microwave field. We chose
    $W_{12}=W_{21}= 0.5g^2\o_0$,
    $\Lambda^{(3)}= 0.6\,g^3\o_0\sim 0.002$, $\O=1.3\o_0$, $M_T = 0.1\o_0$, and
    $M_L=0$.}
\label{transv}
\end{center}
\end{figure}
\begin{center}
\begin{figure}
\includegraphics[scale=0.75]{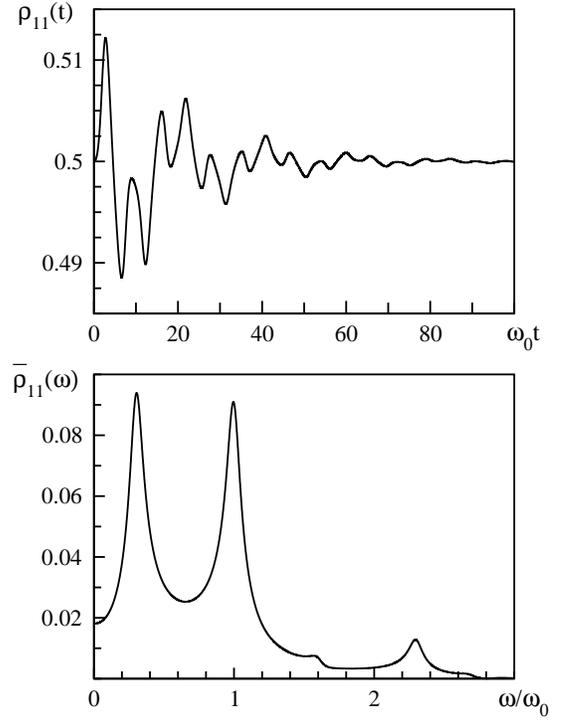}
\caption{Ground state population as a function of time (upper panel)
    and its Fourier transform (lower panel) calculated in presence of
    a strong longitudinal microwave field. The parameters are
    $W_{12}=W_{21}=0.5\,g^2\o_0$, $\Lambda^{(3)}= 0.3\,g^3\o_0\sim 0.001$,
    $\O=1.3\,\o_0$, $M_L = 0.8\,\o_0$, and $M_T=0$.}\label{long}
\end{figure}
\end{center}

Due to the assumption of transverse coupling to the noise source
and of the frequency independence of the third order correlators,
the different coefficients of the third order relaxation matrix
are given by Eqs.(\ref{w21})-(\ref{lambda3}). Note that setting
the coefficients $D_{knml}$ and the longitudinal microwave
contribution $M_L$ to zero one easily recovers Rabi theory. In
this case, solving the eigenvalue equation, one finds the known
Rabi frequency: $\o_{Rabi}=\sqrt{M_T^2/4+(\O-\o)^2}$.

We first discuss the outcomes of a numerical integration of
Eqs.(\ref{mws1})-(\ref{mws2}). The coupling constant and the
renormalized frequency are the same in all the figures: $\o_R =
\o_0 \, \textrm{,}\,\, g^2 = 0.1$.
\begin{center}
\begin{figure}
\includegraphics[scale=0.75]{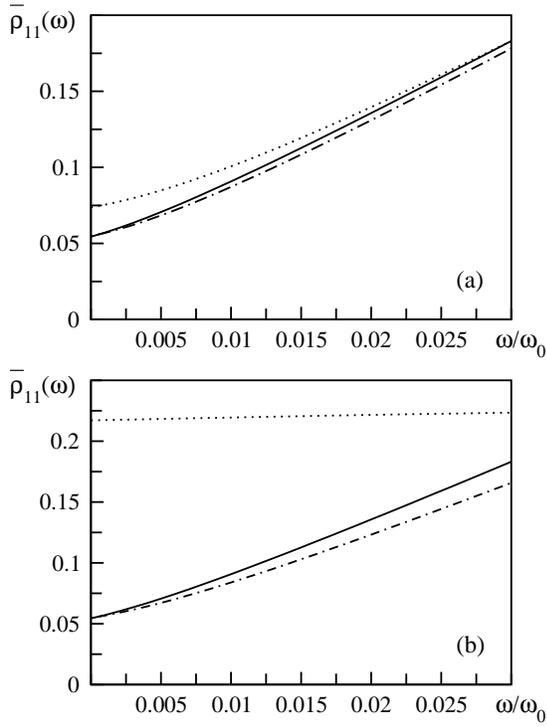}
\caption{Amplitude of the peak at $\omega_0$ as a function of the third order
    correlator $\Lambda^{(3)}$ in absence of microwave field, (solid line), and in
    presence of transverse (dotted line) and longitudinal (dot-dashed
    line) fields. a) Weak microwave fields respectively with $M_T=0.1\,\o_0$ or
    $M_L=0.1\o_0$. b) Strong microwave fields respectively with $M_T=0.8\,\o_0$ or
    $M_L=0.8\o_0$. The other parameters are the same in both the cases:
    $W_{12}+W_{21} = g^2\o_0$, $W_{12}-W_{21} = 0.05\,\o_0g^2$,
    $\O=1.3\,\o_0$.}
\label{cvarsmall}
\end{figure}
\end{center}

\begin{center}
\begin{figure}
\includegraphics[scale=0.75]{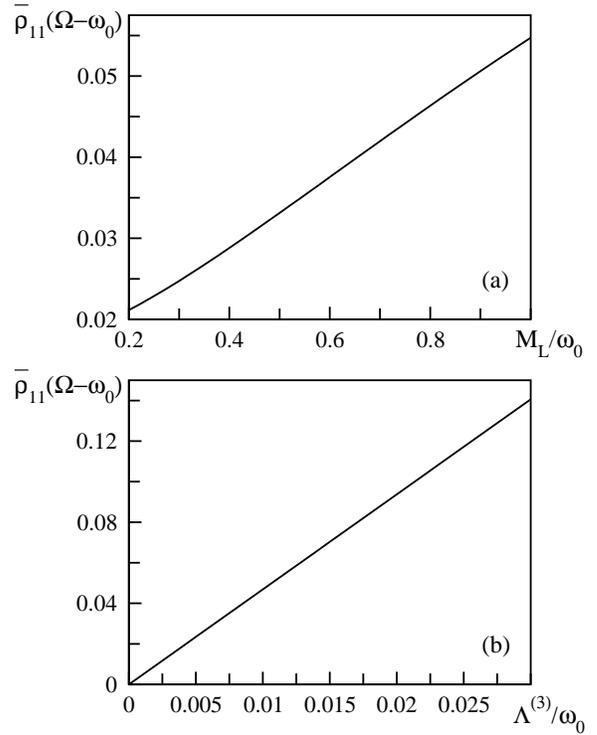}
    \caption{Amplitude of the peak at the detuning frequency $\O-\o_0$
    as a function of of the amplitude of the microwave field $M_L$ (a) and of
    the third order correlator $\Lambda^{(3)}$ (b).
    Values of the parameters: $W_{12}=W_{21}= 0.5\,\o_0g^2$, $\O
    = 1.3\,\o_0$, $M_T=0$. In (a) $M_L=0.8\,\o_0$ while in
    (b) $\Lambda^{(3)}=0.01\,\o_0$.}
\label{detpeak}
\end{figure}
\end{center}
In Fig.\ref{transv} we show  $\r_{11}(t)$ and its Fourier
transform in presence of a weak transverse microwave field. The
structure superimposed to the damped Rabi oscillation can be
better understood by looking at the Fourier transform. We see
indeed four peaks. The zero frequency peak corresponds to a pure
damping term. At the detuning frequency, $\O-\o_0$, we see a large
Rabi peak whose amplitude does not essentially depend on
$\Lambda^{(3)}$ and whose width is fixed by the relaxation rate
$\Gamma=W_{21}+W_{12}$. At frequency $\o_0$ we find the
contribution arising from the third moment fluctuations which is
essentially not modified by the presence of a weak transverse
microwave and that depends on $\Lambda^{(3)}$ linearly. The last
small peak at $\O+\o_0$ does not originate from the third moment
but is due to non-secular terms already present in second order in
the coupling to the
environment.\\

In Fig.\ref{long} we show our results in case of a strong
longitudinal field. In order to suppress the second order effects,
we set $W_{12}=W_{21}$; then, in the absence of third-order
effects, one would simply have a constant ground state population
equal to its asymptotic value $\r_{11}(t)=\r_{\infty}=0.5$. The
non-trivial time-dependence in the ground state population is
therefore completely related to the presence of the third moment
fluctuations. In the case of longitudinal field in Fig.\ref{long}
(lower panel), there are two peaks of non-negligible amplitude in
the Fourier transform at the frequency $\o_0$ and at the detuning
frequency, respectively. The $\o_0$ peak is the peak present also
in the absence of the microwaves (both its amplitude and position
are essentially not affected by the presence of the microwave
field). The second peak, located at frequency $\O-\o_0$, is a
combined effect of the third moment fluctuations and of the
microwave field; its amplitude is directly proportional both to
the value of $\Lambda^{(3)}$ and of $M_L$. The position of this
peak is determined solely by the detuning frequency and it is not
affected by the amplitude of the microwave field, $M_L$.

In the case of pure longitudinal field an approximate analytical
solution of  Eqs.(\ref{mws1})-(\ref{mws2}) can be found. Up to
third order in the coupling constant, $g$, we obtain \bea
    \r_{11}(t)&=& a+be^{-(W_{12}+W_{21})t}+
    \nonumber\\
    & & +
    c\sum_{k=-\infty}^{\infty}J_k\!\lf(\frac{M_L}{\O}\rg)
    \sin\lf(\lf(\o_0-k\O\rg)t+\theta_k\rg)  \,\, .
    \nonumber\\
\label{detsol}\eea In the previous equation $J_k(z)$ is the k-th
Bessel function~\cite{gs}, the phases $\theta_k$ and the real
constants $a$, $b$ and $c$ are defined as \bea
    \theta_k&
    =&\arctan\lf(\frac{k\O-\o_0}{W_{12}+W_{21}}\rg)
    \\
    a&=&\frac{W_{21}}{{W_{12}+W_{21}}}-\frac{\Lambda^{(3)}}{4\O}\!
    \sum_{k=-\infty}^{\infty}\!J_k\!\lf(\frac{M_L}{\O}\rg)
    \sin\lf(\theta_k\rg)\\
    b&=&\hf-\frac{W_{21}}{W_{12}+W_{21}}\\
    c&=&\frac{\Lambda^{(3)}}{4\O}
\eea The peak at $\o_0$ is associated to the $k=0$ contribution of
the sum while the detuning frequency peak and the peak at
frequency $\O+\o_0$ are related respectively to the $k=\pm 1$
-terms. The small contribution connected with the $k=2$ Bessel
function is also visible in Fig.\ref{long} at frequency
$\o=2\O-\o_0$.

\begin{center}
\begin{figure}
\includegraphics[height=5cm,width=7.8cm]{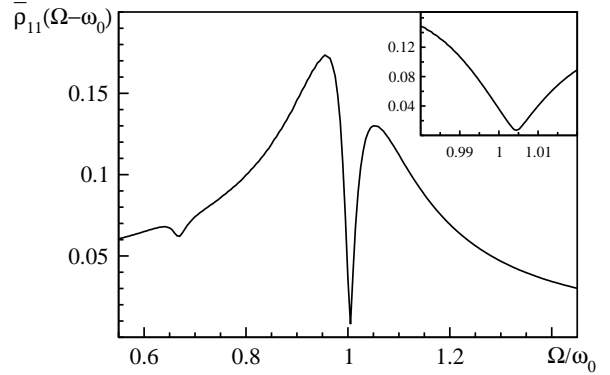}
\caption{Amplitude of the peak at the detuning frequency as a
function of the
    $\O-\o_0$ under longitudinal microwave coupling.
    Values of the parameters: $W_{12}=W_{21}= g^2\,\o_0$, $M_T=0$,
    $M_L=0.8\,\o_0$, $\Lambda^{(3)}=0.01\,\o_0$. The inset is a magnification in the
    region $\Omega \simeq \omega_0$.}
\label{det}\end{figure}\end{center}

 In the hope to use the method discussed in this
paper for the diagnostics of the third-moment of current
fluctuations it is useful to analyze the amplitudes of the
different peaks in some detail. In Fig.\ref{cvarsmall} the height
of the peak at $\o_0$ is shown as a function of $\Lambda^{(3)}$.
In Fig.\ref{cvarsmall}a we show the results in case of weak
fields. In this case, the presence of the longitudinal or the
transverse field does not affect the height and the position of
the third order peak. In Figure \ref{cvarsmall}b we display the
results in case of strong fields. As one could expect from
Eqs.(\ref{mws1})-(\ref{mws2}), a strong transverse field masks
completely the third order dependence of the $\o_0$ peak; on the
other hand, even a strong longitudinal microwave field does not
essentially  modify the height and the position of the third order
peak at frequency $\o_0$. Note that the range $[0,0.03]$ for
$\Lambda^{(3)}/\o_0$ is chosen so that the ratio between second
and third
order contribution varies between 0 and g.\\
Figures \ref{detpeak} and \ref{det}  are devoted to the study of
the amplitude of the peak at the detuning frequency, $\O-\o_0$, in
case of pure longitudinal field. In order to compare the numerical
results shown in these figures with the approximate analytical
solution (\ref{detsol}), we now give the explicit expression of
the $k=1$ term, which is responsible for the peak at the detuning
frequency. This term can be rewritten as \bea
    & & c
    J_{1}\!\lf(\frac{M_L}{\O}\rg) \sin\lf(\Delta
    t+\theta_{1}\rg)\simeq\nonumber\\& &\frac{\Lambda^{(3)} M_L}{8
    \O}\frac{(W_{12}+W_{21})\sin\lf(\Delta t\rg)-\Delta\cos(\Delta
    t)}{(W_{12}+W_{21})^2+\Delta^2}
\label{df}\eea where we set $\Delta=\O-\o_0$ and in the last step
we kept only the linear term in the field amplitude. In
Fig.\ref{detpeak}(a) and Fig.\ref{detpeak}(b) we display the
amplitude of the peak, respectively, as a function of
$\Lambda^{(3)}$ and of $M_L$. As one could expect, the amplitude
of the peak is proportional to $\Lambda^{(3)}$; moreover, as one
can see in Fig.\ref{detpeak}b, the linear approximation is
fulfilled also in case of strong fields. In Fig.\ref{det}, we show
the amplitude of the peak as a function of the detuning frequency;
again the result is as expected based on Eq.(\ref{df}).

\section{Experimental considerations}
\label{exps} Starting from the results presented in the previous
sections, we would like to discuss an experimental protocol for
the measurement of third order noise using a two-level probe
quantum system. The experimental realization of this kind of
measurements is rather delicate. The basic reasons for this is
that a set of inequalities has to be satisfied. First, the level
spacing has to exceed the temperature in the experiment, $T$, in
order to avoid thermal excitations: $\hbar\omega_0 \gg k_BT$.
Second, the time resolution in the experiment, $\delta t$, has to
be good enough to follow the detuned coherent oscillations at
angular frequency $\Omega -\omega_0$, i.e., $\delta t \ll |\Omega
-\omega_0|^{-1}$. Yet the oscillation in the ground state
population has to be measurable, which means that $|(\Omega
-\omega_0)/\omega_0|$ should not be too close to zero. This
condition is determined by the resolution in measuring the
population variations: for very small $|(\Omega
-\omega_0)/\omega_0|$, $\tilde{P}_0(\Omega -\omega_0)$ is
significantly suppressed as demonstrated in Fig. \ref{det}.
Collecting these conditions we have
\begin{equation} \label{ineqs}
|(\Omega -\omega_0)/\omega_0| \ll \frac{1}{\omega_0 \delta t} \ll
\frac{\hbar}{k_BT\delta t}.
\end{equation}
As to a concrete realization, one may employ a hysteretic DC-SQUID
in the tunnelling regime~\cite{claudon04}. The strength of the
method lies in the high contrast in resolving level occupations:
tunnelling rate from the excited state is typically two to three
orders higher from the excited state as compared to that from the
ground state. Measurement of this decay is straightforward by
observing the switching statistics, i.e., the measurement is
typically repetitive in nature. Occupation probabilities of order
0.1 or even below are measurable with adiabatic detection pulses
of $\delta t \sim 1$ ns duration. Measurements are typically
carried out at $T\simeq 30$ mK. Taking this temperature, the
condition at the right end of Eq. (\ref{ineqs}) then states that
$\omega_0 \gg 3\cdot10^9$ s$^{-1}$. Typical level separations
(plasma frequencies) of Josephson junctions are in the range 1 GHz
$< \omega_0/2\pi <$ 100 GHz; thus these values are compatible with
the operation temperature. With $\omega_0\sim 10^{10}$ s$^{-1}$
and $\delta t = 1$ ns, we match the frequency vs. temperature
condition with a margin of factor three. The other critical
condition in Eq. (\ref{ineqs}), $|(\Omega -\omega_0)/\omega_0| \ll
(\omega_0 \delta t)^{-1}$, can then be matched by requesting
$|(\Omega -\omega_0)/\omega_0| \ll 0.1$. Since the maximum of
$\tilde{P}_0$ is obtained at $|\Omega/\omega_0-1| \simeq 0.05$, we
notice that both the inequalities can be satisfied, although
barely.\\
\begin{figure}
\begin{center}
\includegraphics{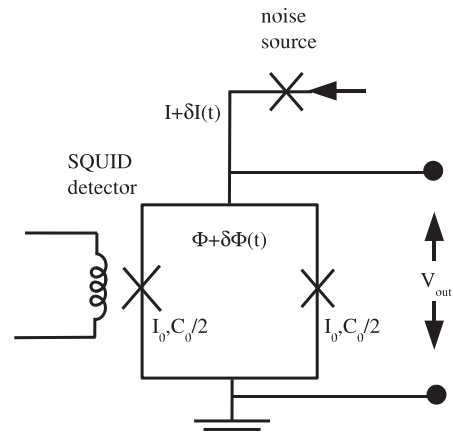}
\caption{Schematic diagram of a possible
    experimental setup to detect third order fluctuations}
\label{app}
\end{center}
\end{figure}
The remaining questions then concern the coupling of the noise to
the detector. As we have already pointed out,  in order to see
oscillations in the occupation probability  due to purely third
order effects, one needs to couple the noise source to $\sigma_x$
and the microwaves to $\sigma_z$. In case of a DC-SQUID detector,
this means that one should couple the noise source through the
current and the external microwave field through the flux. A
schematic diagram of a possible measuring apparatus is shown in
Fig.\ref{app}. The detector is constituted by a DC-SQUID of
negligible inductance formed with two identical Josephson
junctions of critical current $I_0$ and of capacitance $C_0/2$
biased by external flux, $\Phi$, and current, $I$. The
non-equilibrium noise source, which can be a tunnel junction or
another nanostructure, induces time dependent fluctuations in the
biasing current, $\d I(t)$. Finally, the external microwave field
is coupled inductively to the SQUID ring and leads to
monochromatic flux fluctuations, $\d \Phi(t)$. The effective two
level hamiltonian of this system including flux and current
fluctuations can be written as follows \be \label{h}
    H=-\frac{\hbar \o_p}{2}\s_z-\hbar \o_p\lf[N_T(t)
    \s_x -N_D(t)\s_z\rg]
\ee The operators  $N_T(t)$ and $N_D(t)$ are defined by the
following equations \bea
    N_T(t)& =&t_I\frac{\d
    I(t)}{I_0}+\lf(t_I-2\l d_\F\rg)\frac{\pi \d \F(t)}{\F_0}\\
    N_D(t)&=&\lf(d_{\F}+2\l t_\F\rg)\frac{\pi\d
    \F(t)}{\F_0} +2\l t_I\frac{\d
    I(t)}{I_0}
\eea
where $\F_0$ is the elementary flux quantum.
In the previous equations we denoted with $\l$ and $\o_p$, respectively,
the SQUID anharmonicity  and the SQUID plasma frequency:
\be
    \l=\frac{\eta}{6}\frac{I}{I_c}\lf(1-\frac{I^2}
    {I_c^2}\rg)^{-\frac{5}{8}}\quad\o_p^2=\frac{2\pi}{\F_0C_0}\sqrt{I_c^2-I^2}
\ee
with $I_c=2I_0\cos \frac{\pi\Phi}{\F_0}$ and
$\eta=\hbar^{\hf}(C_0I_c)^{-\frac{1}{4}}\lf(\frac{\F_0}{2\pi}\rg)^{-\frac{3}{4}}$.\\
Moreover we have introduced the adimensional parameters $t_I,\, t_\Phi,\, d_\F$:
\be
\label{par}
    t_I=\frac{1}{\eta}\frac{I_0}{I_c}
    \lf(1-\frac{I^2}{I_c^2}\rg)^{-\frac{3}{8}}\!\!\!\!\!\!,\,\,
    t_{\F}=\frac{I}{I_0}t_I\tan \frac{\pi\F}{\F_0},\,\,
    d_\F=\tan \frac{\pi\F}{\F_0}\,.
\ee

As it is clear from Eqs.(\ref{h})-(\ref{par}) a  transverse
coupling to current fluctuations and a longitudinal coupling to
flux fluctuations (i.e. to the microwave field), can be
simultaneously realized only in case of zero or very small DC
component of the external biasing current. Experimentally this
condition could be obtained subtracting the DC component of $I(t)$
by means of a superconducting line, as it was shown in
Ref.\onlinecite{pekola2}. In this case the SQUID potential becomes
harmonic and one should also take into account the possibility to
have transitions to higher levels. Anyway, if initially only the
first two levels are occupied, this effect is of  order higher
than the third in the system-noise coupling $g$. \\
Assuming that the transverse coupling condition is fulfilled, that
is  $I<<I_c$, we now evaluate the amplitude of the third order
oscillations for the detector's scheme presented in
Fig.\ref{app}.\\
Comparing Eq.(\ref{h})-(\ref{par}) to the model hamiltonian in
Eq.(\ref{heff}) and using the definition given in equation
(\ref{lambda3}), we rewrite  $\Lambda^{(3)}$ as follows
\bea\label{l3ord}\Lambda^{(3)}&\!\! =&\!\! \lf(\frac{\o_p}{\eta
I_c}\rg)^3\!\!\!\int\! dt_1\!\int\! dt_2\Big[\lf<\d I(t_1) T
\lf(\d I(t_2)\d I(0)\rg)\rg>+\nonumber\\& & +\frac{1}{3}\lf<\tilde
T \lf(\d I(t_1)\d I(t_2)\d I(0)\rg)\rg>\Big]. \eea Using the
results derived by Salo, Hekking and Pekola in
Ref.\onlinecite{salo} within the framework of scattering
theory\cite{blanter99}, we can easily obtain an explicit
expression of $\Lambda^{(3)}$ in terms of the transmission
eigenvalues, $T_n$,
and of the voltage bias, $V$, across the junction.\\
In particular, in case of energy independent scattering in the
limit of zero temperature of the noise source, we have
\be\label{l3fin}\Lambda^{(3)}=\frac{4}{3}\lf(\frac{e\o_p}{\eta
I_c}\rg)^3
 \frac{eV}{h}\sum_n\,T_n(1-T_n)(1-2T_n)\, ;\ee
 as one can see, in this limit,
 $\Lambda^{(3)}$ is proportional to the
usual third cumulant of current
statistics\cite{levitov96,levitov04}. Eventually, using
Eq.(\ref{l3fin}), we can also check the validity of the
perturbative hypothesis and give a rough estimate of the ratio
between the third and the second order contributions to the qubit
dynamics. In the limit of zero frequency and zero temperature we
have:
\be\label{l3w21}\frac{\Lambda^{(3)}}{W_{21}}\sim\frac{2}{3}\tilde
g\frac{F_3}{F_2}\ee  where $\tilde g=e\o_p/(\eta I_c)\simeq
(C_0\,I_c)^{-\frac{1}{4}}\cdot(2.1\,\,\,10^{-6}
(A\cdot\textrm{farad})^{\frac{1}{4}}) $ is an adimensional
coupling constant and  $F_2$ and $F_3$ are the Fano factors of the
second and of the third order: $ F_2=\sum_n T_n(1-T_n)/\sum_n T_n
$ and $ F_3=\sum_n T_n(1-T_n)(1-2T_n)/\sum_n T_n $. In deriving
Eq. (\ref{l3w21}) the well-known relation between the second order
transition amplitude and the the Fano factor $F_2$ has been used,
see, for example, Refs.\onlinecite{deJong97,aguado}.
\section{Conclusions}
\label{conclusions}

In this paper we analyzed the possibility to use solid state
qubits as detectors for higher moments of current fluctuations. We
showed that in some cases there are distinct features, due to the
non-secular terms in the Master equation, solely related to the
presence of the third moment of current fluctuations. This may be
a very interesting circumstance as usually these additional
effects are masked by the large background coming from the noise
(second order cumulant in the fluctuations). After having derived
the general form of the Master equation up to the third order in
the coupling between the environment and the bath we considered in
some detail a two-level system coupled to a noise source. Indeed
we found that, in the presence of purely transverse noise, the
population in the ground state oscillates at a frequency
$\omega_0$ if the two-level system is initially prepared in a
superposition. The difficulty of measuring these high frequency
oscillations can be alleviated by applying a microwave field. In
this case the oscillations associated to the third-moment are
pushed down to the detuning frequency $\Omega -\omega_0$.

A possible experimental implementation of this scheme of detection
has been discussed in Section~\ref{exps}. As a two-level system
(the detector) we considered a DC-SQUID and discussed the range of
applicability of the scheme. Combining the narrow margins in
experimental parameters and the rather unfavourable coupling of
noise to the detector, it is obvious that measurement of the
effects predicted here is not straightforward using a DC-SQUID as
a sensor. It remains to be analyzed if other controllable
two-level systems (charge qubits for example) may be more suited
as detectors. Nevertheless we find interesting the existence of
features entirely due to the higher moments of current
fluctuations.

\acknowledgments We would like to thank G. Falci, T. Heikkil\" a,
T. Ojanen and F. Taddei for fruitful discussions. Financial
support from EU (grants SQUBIT2, EUROSQIP, RTNANO), IUR (grant
Prin 2005) and Institut Universitaire de France is acknowledged.


\begin{thebibliography}{99}
\bibitem{deJong97}
        M. J. M. de Jong, and C. W. J. Beenakker, in {\em
        Mesoscopic Electron Transport}, edited by L. L. Sohn, L. P.
        Kouwenhoven, and G. Sch\"on (Kluwer Academic Publishers, Dordrecht,
        1997).
\bibitem{blanter99}
        Ya. M. Blanter, M. Buttiker, Phys. Rep. {\bf 336}, 1 (2000).
\bibitem{Kogan96}
        S. Kogan, {\it Electronic Noise and Fluctuations in Solids}, Cambridge
        University Press, Cambridge, 1996.
\bibitem{depicciotto97}
        R. de Picciotto, M. Reznikov, M. Heiblum, V. Umansky, G. Bunin,
        and D. Mahalu, Nature {\bf 389}, 162 (1997).
\bibitem{saminadayar97}
        L. Saminadayar, D. C. Glattli, Y. Jin, and B. Etienne,
        Phys. Rev. Lett {\bf 79}, 2526 (1997).
\bibitem{delft}
        {\sl Quantum Noise in Mesoscopic Physics}, edited by Y.V. Nazarov,
        NATO Science Series in Mathematics, Physics and Chemistry (Kluwer,
        Dordrecht, 2003).
\bibitem{levitov96}
        L.S. Levitov, H.B. Lee, and G.B. Lesovik, J. Math. Phys. {\bf 37}, 4845 (1996).
\bibitem{reulet03}
        B. Reulet, J. Senzier, and D. E. Prober, Phys. Rev. Lett. {\bf 91}, 196601
        (2003).
\bibitem{beenakker03}
        C.W.J. Beenakker, M. Kindermann, and Yu.V. Nazarov
        Phys. Rev. Lett. {\bf 90}, 176802 (2003).
\bibitem{bomze05}
        Yu. Bomze, G. Gershon, D. Shovkun, L. S. Levitov, and M.
        Reznikov,
        Phys. Rev. Lett. {\bf 95}, 176601 (2005).
\bibitem{lindell04}
        R.K. Lindell, J. Delahaye, M. A. Sillanp\"a\"a, T. T. Heikkil\"a,
        E. B. Sonin, and P. J. Hakonen, Phys. Rev. Lett. {\bf 93}, 197002 (2004).
\bibitem{lesovik94} G.B. Lesovik,  JETP Lett., {\bf 60}, 820 (1994)

\bibitem{aguado}
    R. Aguado and L.P. Kouwenhoven, Phys. Rev. Lett. {\bf 84}, 1986 (2000).
\bibitem{heikkila04}
        T.T. Heikkil\"a, P. Virtanen, G. Johansson,
        and F.K. Wilhelm, Phys. Rev. Lett. {\bf 93}, 247005 (2004).
\bibitem{ankerhold05}
        J. Ankerhold, H. Grabert, Phys. Rev. Lett. {\bf 95}, 186601 (2005).
\bibitem{tobinska04}
        J. Tobiska and Yu.V. Nazarov, Phys. Rev. Lett. {\bf 93}, 106801 (2004).
\bibitem{lesovik06}
G.B. Lesovik, F. Hassler and  G. Blatter, Phys. Rev. Lett. {\bf
96}, 106801 (2006).
\bibitem{pekola04}
        J.P. Pekola, Phys. Rev. Lett. {\bf 93}, 206601 (2004).
\bibitem{devoret00}
        M.~H. Devoret and R.~J. Schoelkopf, Nature {\bf 406},  1039  (2000).
\bibitem{astafiev04}
        O. Astafiev, Yu. A. Pashkin, Y. Nakamura, T. Yamamoto, and J. S. Tsai
Phys. Rev. Lett. {\bf 93}, 267007 (2004)
\bibitem{ojanen05}
        T. Ojanen and T. T. Heikkil\"a, Phys. Rev. B {\bf 73}, 020501(R) (2006).
\bibitem{falci}
        Recently the effect of counter-rotating terms has been discussed for
        the dynamics of a Josephson qubit in: A. D'Arrigo, G. Falci, A. Mastellone, E.
        Paladino, Physica E {\bf 29}, 297 (2005).
\bibitem{claudon04}
        J. Claudon, F. Balestro, F.W.J. Hekking, and O. Buisson, Phys.
        Rev. Lett. {\bf 93}, 187003 (2004).
\bibitem{books}
        H.~J. Carmichael, {\em An Open Systems Approach to Quantum Optics}
        (Springer-Verlag, Berlin, 1993);
        K. Blum, {\em Density Matrix Theory and Applications}, 2 ed.
        (Plenum Press, New York, 1996);
        C. Cohen-Tannoudji, J. Dupont-Roc, and G. Grynberg,
        \emph{Atom-Photon interactions}, (Wiley, New York, 1992).
\bibitem{gs}
    I.S. Gradshteyn, I.M. Ryzhik, \emph{Table of Integrals, Series and
    Products} (Academic Press, New York, 1980).
\bibitem{pekola2}
    J. P. Pekola, T. E. Nieminen, M. Meschke, J. M. Kivioja,
    A. O. Niskanen, and J. J. Vartiainen,
        Phys. Rev. Lett. {\bf 95}, 197004 (2005).
 \bibitem{salo} J. Salo, F.W.J. Hekking, J.P. Pekola,
 cond-mat/0605478, (2006)
 \bibitem{levitov04} L.S. Levitov and M. Reznikov, Phys. Rev. B 70, 115305 (2004).
\end{thebibliography}
\end{document}